\RequirePackage{lineno}  
\documentclass[preprint,12pt]{elsarticle}
\usepackage{dcolumn}
\usepackage{bm}
\usepackage{graphicx}
\usepackage[dvips]{color}
\usepackage{epsfig}
\usepackage{float}
\usepackage{subfig}
\usepackage{setspace}
\usepackage{rotating} 

\usepackage{amssymb}

\biboptions{compress}

\journal{NIMA Proceedings}

\begin{document}

\begin{frontmatter}



\title{Astroparticle Physics with a Customized Low-Background Broad Energy Germanium Detector}



\author[pnnl]{C.E.~Aalseth}
\author[lbnl]{M.~Amman}
\author[usc,ornl]{F.T.~Avignone III}
\author[ncsu,tunl]{H.O.~Back}
\author[itep]{A.S.~Barabash}
\author[chi]{P.S.~Barbeau}
\author[lbnl]{M.~Bergevin}
\author[ornl]{F.E.~Bertrand}
\author[lanl]{M.~Boswell}
\author[jinr]{V.~Brudanin}

% ...
\address[itep]{Institute for Theoretical and Experimental Physics, Moscow, Russia}
\address[jinr]{Joint Institute for Nuclear Research, Dubna, Russia}
\address[lanl]{Los Alamos National Laboratory, Los Alamos, NM, USA}
\address[lbnl]{Lawrence Berkeley National Laboratory, Berkeley, CA, USA}
\address[ncsu]{Department of Physics, North Carolina State University, Raleigh, NC, USA}
\address[ornl]{Oak Ridge National Laboratory, Oak Ridge, TN, USA}
% ...
\address[pnnl]{Pacific Northwest National Laboratory, Richland, WA, USA}
\address[tunl]{Triangle Universities Nuclear Laboratory, Durham, NC, USA}
% ...
\address[chi]{Department of Physics, University of Chicago, Chicago, IL, USA}
% ...
\address[usc]{Department of Physics and Astronomy, University of South Carolina, Columbia, SC, USA}
% ...


\begin{abstract}
The {\sc Majorana} Collaboration is building the {\sc Majorana Demonstrator}, a 60 kg array of high purity germanium detectors housed in an ultra-low background shield at the Sanford Underground Laboratory in Lead, SD.  The {\sc Majorana Demonstrator} will search for neutrinoless double-beta decay of $^{76}$Ge while demonstrating the feasibility of a tonne-scale experiment.  It may also carry out a dark matter search in the 1-10 GeV/c$^2$ mass range.  We have found that customized Broad Energy Germanium (BEGe) detectors produced by Canberra have several desirable features for a neutrinoless double-beta decay experiment, including low electronic noise, excellent pulse shape analysis capabilities, and simple fabrication. We have deployed a customized BEGe, the {\sc Majorana} Low-Background BEGe at Kimballton (MALBEK), in a low-background cryostat and shield at the Kimballton Underground Research Facility in Virginia. This paper will focus on the detector characteristics and measurements that can be performed with such a radiation detector in a low-background environment.
\end{abstract}


\end{frontmatter}


\section{Introduction}
\label{sec:intro}
Determining the nature of the neutrino would provide insight into fundamental questions about the nature of matter.  Recent results from atmospheric, solar, and reactor-based neutrino oscillation experiments (Super-Kamikande, KamLAND, and SNO) \cite{Ash04,Ara05,Ahm04} have shown that neutrinos have mass and oscillate.  This is a clear indication that the standard model of nuclear physics is incomplete.  Understanding the neutrino character (Majorana or Dirac), the neutrino mass spectrum, the neutrino mass generation mechanism, and the absolute mass scale of the neutrino are all questions that will be addressed by future experiments.

The {\sc Majorana Demonstrator} is a mixed array of enriched (86\%) and natural germanium crystal detectors that will search for neutrinoless double beta decay ($0\nu\beta\beta$) of $^{76}$Ge. The discovery of this decay would signal that the neutrino is a Majorana fermion (its own anti-particle) and that lepton number is violated, having significant implications for our understanding of the nature of the neutrino and fundamental interactions. It may also perform a search for Weakly Interacting Massive Particles (WIMPs), a promising candidate for dark matter, by searching for nuclear recoils of the germanium nucleus.

\section{P-Type Point Contact Detectors}
\label{sec:ppc}
P-type point contact (P-PC) Ge detectors \cite{Luk89,Bar07}, a relatively new development in semiconductor detector technology, have been demonstrated to provide both exceptional energy resolution ($\sim$150 eV at 10.36 keV) and low-energy thresholds ($\sim$400 eV) \cite{Bar07,Aal08,Aal10}. They allow multi-site event discrimination via pulse shape analysis that is comparable or superior to discrimination attained using highly segmented coaxial type detectors \cite{Bud09,Coo10b}. Such low-energy performance is required for dark matter searches, and provides enhanced background rejection for certain $0\nu\beta\beta$ backgrounds. Numerous successful prototypes have been commercially produced and successfully operated \cite{Bar07,Aal08}.  P-PC-like detectors, commercially available from Canberra as ÒBroad Energy GermaniumÓ (BEGe) detectors, have performed robustly in multiple low-background (LB) underground deployments \cite{Aal10}. We have chosen a slightly modified version of the P-PC BEGe detectors to be deployed in the {\sc Majorana Demonstrator}.

\subsection{Pulse Shape Discrimination with P-PCs}
\label{sec:psd}
The pulse shape discrimination (PSD) method {\sc Majorana} will implement is outlined in \cite{Coo10b}.  The P-PC geometry yields excellent charge collection and signal induction properties which fit the needs of a multi-site PSD. Figure~\ref{fig:Figure1} shows the drift properties for a prototype BEGe detector. It shows the weighting potential (WP) \cite{Sch38} associated with the point contact in addition to the charge carrier (holes) drift paths associated with interactions within the crystal.  The figure shows that the WP falls off rapidly and nearly isotropically. This results in long charge drift times followed by the rapid charge collection around the point contact.  

\section{MALBEK - {\sc Majorana} Low-Background BEGe at Kimballton}
\label{sec:malbek}
\begin{figure}[htbp]
\centering
\includegraphics[width=0.7\textwidth]{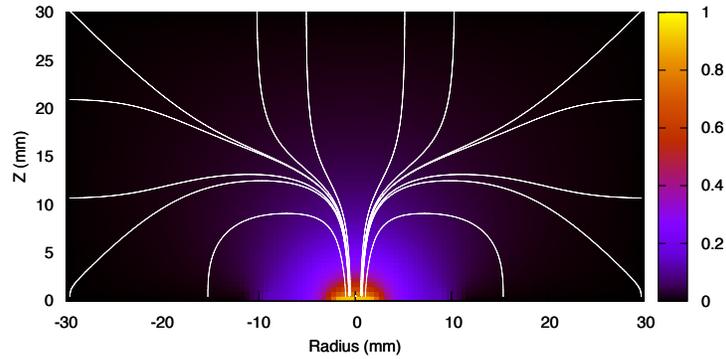}
\caption{Weighting potential (WP) and charge carrier (holes) drift paths; WP quickly falls off nearly isotropically, resulting in long charge drift times followed by the rapid charge collection around the point contact.  This results  in distinct current pulses for each interaction. \cite{Coo10a}.}
\label{fig:Figure1}
\end{figure}

As part of the research and development efforts for the {\sc Majorana Demonstrator} we have deployed MALBEK, a prototype BEGe detector, at the Kimballton Underground Research Facility (KURF) in Ripplemead, VA.  KURF is located at a depth of 1450 meters water equivalent (m.w.e.).  MALBEK is a 450 g natural Ge crystal housed in a LB cryostat and outfitted with custom made LB components.  MALBEK differs from commercial BEGe detectors in three respects:  (1) absence of typical thin front entrance window, (2) the passivation ditch radius is almost a factor of two larger (15 mm), and (3) the point contact size is nearly a factor of two smaller (3.5 mm).  The reduction of the point contact should decrease the capacitance, and hence the electronic noise.  Simulations (see Figure~\ref{fig:Figure1}) indicate the larger ditch radius should optimize the charge collection and depletion regions within the crystal \cite{Coo10a}.  MALBEK is housed in a LB copper cryostat with custom LB internal components.  The lead shielding is comprised of one inch of ancient lead and 8 inches of standard lead.  The entire lead shield and internal cavity is continuously purged with liquid nitrogen boil off to reduce potential radon backgrounds.  Additional shielding is provided by 10 inches of polyethylene and a muon veto (not used in the current analysis).

\subsection{Detector Characteristics}
\begin{figure}[htbp]
\centering
\includegraphics[width=0.7\textwidth]{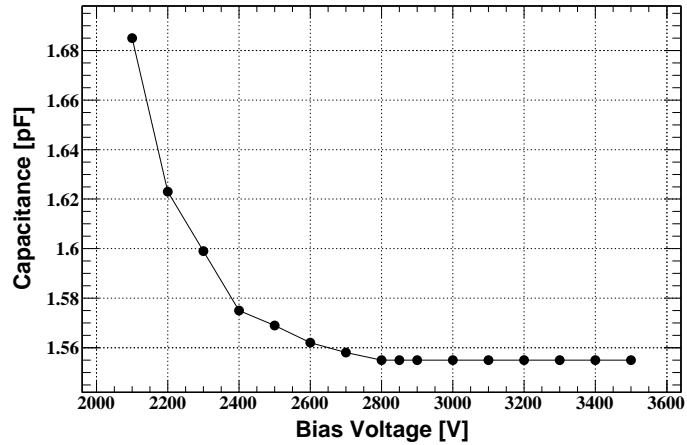}
\caption{Detector capacitance vs. bias voltage, 1.55 pF at depletion.}
\label{fig:Figure2}
\end{figure}

\begin{figure}[htbp]
\centering
\includegraphics[width=0.7\textwidth]{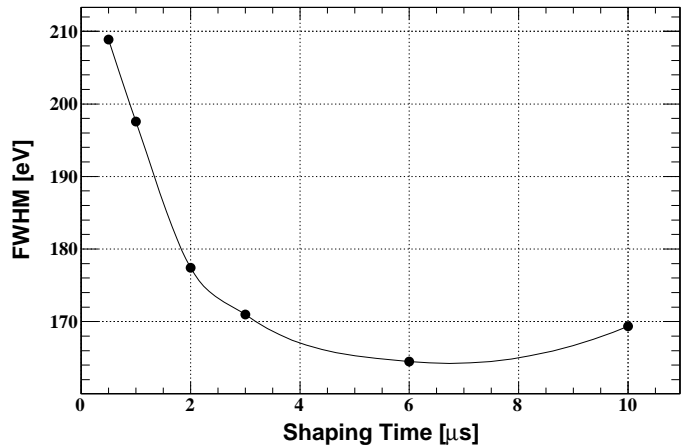}
\caption{Electronic noise of MALBEK as measured by a pulser, noise corner located at 6 $\mu$s, $\sim$165 eV.}
\label{fig:Figure3}
\end{figure}

The physical dimensions of the crystal are 60 mm $\times$ 30 mm (diameter $\times$ height), with a 15 mm ditch radius.  The capacitance of MALBEK has been measured to be $\sim$1.55 pF, as shown in Figure~\ref{fig:Figure2}.  The noise corner is located at 6 $\mu$s, $\sim$165 eV, as shown in Figure~\ref{fig:Figure3}.

\subsection{Data Acquisition}
\label{sec:daq}
The design of the MALBEK data acquisition system (DAQ) is driven by several experimental goals and requirements.  First, the KURF underground environment necessitates a DAQ that is highly automated and remotely controllable.  Second, the wide energy range of interest requires a DAQ capable of triggering on pulses ranging from several hundred eV to greater than 2000 keV.  Lastly, the MALBEK experiment provides a test bed for electronics that may be used with the {\sc Majorana Demonstrator}, requiring the DAQ to be modular and easily reconfigurable.

Analog signals from the MALBEK detector are amplified by a custom Canberra low-noise pulse-reset charge sensitive preamplifier.  The preamplifier has two amplified signal outputs.  One of these is digitized by a Struck Innovative Systeme 3302 ADC.  The SIS3302 is a VME based, 8 channel, 16-bit digitizer capable of digitization rates up to 100 MHz.  The SIS3302 incorporates a fast FIR trapezoidal triggering filter capable of triggering on low amplitude signals and allows for data readout in parallel with acquisition.  

The MALBEK data acquisition system uses the Object-oriented Real-time Control and Acquisition (ORCA) system, an object oriented data acquisition application that provides an easy to use, graphical interface for manipulation of experimental hardware and data streams \cite{How04}.  ORCA is self-monitoring, sending email notification and alarms to operators based on user configurable preferences.  The implemented MALBEK DAQ is currently operating underground in a production data taking mode and is remotely accessible 24 hours a day.

\subsection{Data Analysis}
\label{sec:analysis}
\begin{figure}[htbp]
\centering
\includegraphics[width=0.7\textwidth]{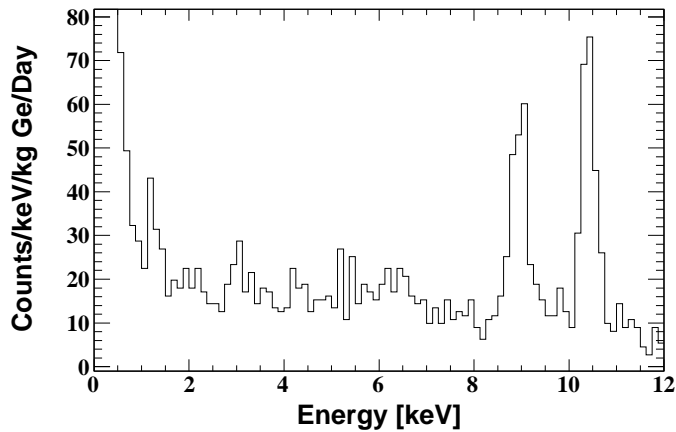}
\caption{20 day spectrum for MALBEK with no cuts (runs in which liquid nitrogen (LN) filled are not used).  Dominant features include cosmogenic peaks from K-shell EC in $^{65}$Zn (8.99 keV) and $^{68,71}$Ge (10.36 keV) and L-shell peaks from $^{68,71}$Ge (1.1, 1.29 keV).  Noise pedestal located at $\sim$735 eV. Even with only the LN fills removed, the continuum is within a factor of two of \citet{Aal08,Aal10}.  Further improvements are expected with more comprehensive cuts.}
\label{fig:Figure4}
\end{figure}

Analysis of digitized waveforms allows the extraction of useful waveform characteristics, such as rise-time, baseline, peak height, extrema values, derivative, integral, etc.  Off-line event-by-event processing also enables further digital processing of the waveform, for example pulse smoothing either by wavelet de-noising or performing a moving average.  The MALBEK data analysis framework is as follows:  (1)  remove baseline, (2) smoothing (by one of the techniques listed above), (3) pole zero correction, and (4) extract waveform characteristics.  A 20 day spectrum with no cuts (runs in which liquid nitrogen (LN) filled are not used) is shown in Figure~\ref{fig:Figure4}.  The detector threshold is 735 eV, but we expect significant improvement with the addition of cuts to microphonics and other sources of background.

\section{Future Work}
\label{sec:future}
Implementation of cuts to remove microphonic events and pulse-inhibit induced events will decrease the contribution to the continuum.  We will also be commissioning a $\sim$4$\pi$ muon veto and 1.5 inches of borated polyethylene.  We plan to investigate the performance of MALBEK at various operating temperatures.  The temperature of the field effect transistor (FET) is directly proportional to the series noise, therefore the colder the detector, the smaller the contribution from series noise (parallel noise also has a temperature component, but is only relevant for resistive feedback preamplifiers).  Further refinement of the data analysis is also needed.  Several digital shaping and filtering methods, e.g. trapezoidal and cusp, will be examined to further increase resolution and possibly decrease threshold.


\section{Acknowledgements}
This work was sponsored by DOE Grants DE-FG02-97ER41041 and DE-FG02-97ER41033 and the state of North Carolina.



\bibliographystyle{model1-num-names}







\end{document}